# Surface morphology and superconductivity of Nb thin films by biased target ion beam deposition

Salinporn Kittiwatanakul [1], Nattawut Anuniwat [2], Nam Dao [1], Stuart A. Wolf [1,2], Jiwei Lu [1]

[1] Department of Materials Science and Engineering, University of Virginia, Charlottesville VA USA 22904

[2] Department of Physics, University of Virginia, Charlottesville VA USA 22904

**ABSTRACT**

One of many challenges for niobium (Nb) based superconducting devices is the improvement over the surface morphology and superconducting properties as well as the reduction of defects. We employed a novel deposition technique, i.e. biased target ion beam deposition technique (BTIBD) to prepare Nb thin films with controlled crystallinity and surface morphology. We found that the target current ($I_{Target}$) and the target bias ($V_{Target}$) were critical to the crystallinity and surface morphology of Nb films. The high target current ($I_{Target}$ >500 mA and $V_{Target}$ = 400 V bias) during the deposition degraded the Nb crystallinity, and subsequently reduced the critical temperature for superconductivity ($T_c$). $V_{Target}$ was critical to the surface morphology, i.e. grain size and shape and the surface roughness. The optimized growth condition yielded very smooth film with RMS roughness of 0.4 nm that was an order of magnitude smoother than that of Nb films by sputtering process. The critical temperature for superconductivity was also close to the value of the bulk Nb. The quality of Nb film was evident in the presence of a very thin proximity layer (~ 0.8 nm). The experimental results demonstrated that the preparation of smooth Nb films with adequate superconductivity by BTIBD could serve as a base electrode for the in-situ magnetic layer or insulating layer for superconducting electronic devices.

# 1. INTRODUCTION

The preparation of single-crystal like niobium thin films has been a goal of experimentation for many years due to the extensive use of these materials in superconducting devices such as Josephson junctions [1-4]. A recently proposed Josephson Magnetic Random Access Memory (JMRAM) devices require very smooth surface of superconducting Nb electrodes on top of which magnetic multilayer structures are deposited [3-6], and ideally the root-mean-square roughness of Nb thin films should be less than 0.3 nm to ensure the proper magnetic characteristics of magnetic multilayers on top. Previously, the growth and properties of Nb thin films have been extensively explored by various deposition techniques such as sputtering, evaporation, molecular beam epitaxy (MBE), and ion beam deposition (IBD) [7-11]. There has been many studies on the effects of the thickness on the superconductivity of Nb thin films [8,12,13], but few groups explored ways to smooth the surface while preserving the superconducting transition temperature.

The IBD has the advantages of independent control of target and substrate environments, low processing pressures, low sputtering rate, directional sputtered flux, and high adatom flux energy; however it is not suitable for optimum conditions for multilayer deposition, because conventional IBD uses relatively high sputtering ion energies to better focus the ion beam on the target [14,15]. The focused beam is used to prevent the overspill contamination when a large fraction of an unfocused ion beam miss the targets and sputters off other materials from hardware inside the vacuum chamber [16]. The biased target ion beam deposition (BTIBD) technique was developed to improve upon the conventional IBD, with the key advantage of using a low energy ion source [17,18], which produces very high-density inert gas ions with very low energy (5-50eV). The use of the low energy inert ions avoids the overspill contamination problem faced by the conventional IBD system because sputtering only occurred on the biased target. Furthermore, the use of a wide ion beam that improves both target material utilization efficiency and deposition thickness uniformity. The low sputtering energy leads to low kinetic energy of the sputtered material species from the target, which is beneficial to the interfacial smoothness and the reduction of interlayer diffusion mixing. In addition the ion source are designed to be capable of operating over a broader range of processing pressure ($10^{-4}$ to $5\times10^{-3}$ Torr), allowing additional control of the background gas scattering of the inert ions before they reach the substrate. Overall, the BTIBD provides a way of controlling the adatom energies over a very wide range which makes it a well-suited technology for depositing multilayer structures containing sharp interfaces [16–19].

In this paper, the BTIBD technique was used to synthesize Nb thin films on 100 nm $SiO_2$/Si substrates. The effect of growth conditions, e.g. target bias, deposition rate, etc., on surface morphology and crystallinity were studied to investigate the interdependence among microstructure, superconductivity and surface roughness.

## 2. EXPERIMENT

Prior to the deposition, the chamber was pumped down to the base pressure of $5\times10^{-8}$ Torr, during the deposition the operating pressure was $7\times10^{-4}$ Torr. The ion source was started using a flow rate of 60 SCCM of Ar gas and 40 V on the ion source, and a flow rate of 20 SCCM Ar gas and 10 V on the cathode. A Nb target (purity ~99.95%) was used. The cathode current was fixed at 9.5 A for all depositions, while the anode current was adjusted to reach the target current for the desired deposition rate. All of the niobium deposition reported in this work was done without intentional heating, and the sample stage was water cooled to maintain the constant temperature during the deposition.

The main parameters explored in this work included target bias, target current, and film thickness. The film thickness was adjusted by the deposition time while other deposition parameters were fixed. The Nb target was biased negatively (400-900 V) to provide the ion energy necessary for sputtering. The sputtering rate depends on both target bias ($V_{Target}$ or $V_T$) and target current ($I_{Target}$ or $I_T$). The crystallinity of Nb thin film depends on the target current and the film thickness, while the surface roughness highly depends on the target current. When other factors such as crystallinity, deposition rate, and film thickness were ruled out, the grain size and the critical temperature for superconductivity depend on the target bias. The ion flux generated from the ion beam source can be adjusted by anode current ($I_A$) and the Ar flow through anode (ion source), while the ion energy could be raised from the increase of anode voltage. For this study, the target bias was controlled independently, while the target current was adjusted via the anode current with fixed Ar flow through anode in the work presented here.

There were 4 sets of thin film samples deposited on 100 nm amorphous $SiO_2$/ (100) Si substrates via BTIBD with different growth parameters as listed in table I. Sample set A was deposited at a constant target bias of 400 V and a constant target current of 500 mA with thickness range of 10-100 nm; while set B was deposited at a constant target bias of 400 V, but at various target currents (300-650 mA). The samples in set C and D were deposited at various target bias with deposition rates of 0.5 and 0.8 Å/s respectively, with a constant film thickness (~50 nm). In the sets C and D the target current was adjusted such that within each set the deposition rate was constant; while between set C and D, the deposition time was adjusted to get the same Nb thickness (as ex-situ determined by x-ray reflectivity measurements).

X-ray diffractrometry (XRD) (Smartlab[TM] Rigaku Inc.) was used to confirm the phase purity and crystalline orientation of the film ($2\theta$ scans), and to compare the crystallinity of the thin films ($\omega$ scans). The surface morphology of a 1x1 µm$^2$ area was characterized by atomic force microscopy (AFM) (Asylum Research Cypher), and Gwyddion software was used to process the grain analysis. The transport properties of Nb thin films were measured in a Quantum Design PPMS-6000 system using a 4-point method with indium contacts in the temperature range of 4 - 300 K. The superconducting transition temperature ($T_c$) was then extracted from the heat-up and cool-down curves.

## 3. RESULTS AND DISCUSSION

The $2\theta$ scans showed the Nb (110) peak at ~38.068° along with a Si (400) substrate peak for the thickest sample (A5) as seen in figure 1(a), yielding the lattice constant of ~3.34 Å, under a tensile strain (~1.17 %) (bulk value ~3.30 Å). A (110) texture was expected for Nb with Body Centered Cubic (BCC) crystal structure [20]. Figure 1(b) plots $2\theta$ scans around Nb (110) peak for all samples, where the peak positions and the full width half max values (FWHM) of $2\theta$ peak were extracted to calculate the lattice spacing $d_{110}$ and the crystallite size as listed in table I (denoted as crystal.). The rocking curve measurement ($\omega$ scan) was performed on each film [figure 1(c)], and the FWHM values of the ω peaks are summarized in table I. As expected, the lattice spacing $d_{110}$ in set A approached to the bulk value as the film grew thicker.

While in set B, the lattice spacing $d_{110}$ was close to the bulk value when target current used was smaller. The higher target current introduced more tensile strains along the <110>. The strains were found to be increasing from ~1.12% to ~1.71%, as the target current was increased from 300 mA to 645 mA. For sets C and D, the lattice parameter was reduced and approached to the bulk value when the target voltage was decreased. As the target bias was increased from 400 V to 900 V, the tensile strains were 1.12-1.50%, and 1.12-1.45% for set C and set D respectively. These values of strain are in a good agreement of 1.2% expansion of lattice parameter reported on Nb thin films [21]. While the films deposited at similar conditions on 2" wafers show the in-plane stress of -5 to -6 x $10^9$ dyne/cm$^2$ which is comparable to those of other Nb thin films deposited by dc magnetron sputtering [21,22].

AFM images showed two different types of grain shapes: elongated grains [figure 2(b)-(c)] which is a typical grain shape for Nb (110) thin films [11,13], and equiaxed grains [figure 2(a) inset]. The morphology elongated grain shape was associated with better crystallinity as seen in both XRD $2\theta$ scan, and also in FWHM of the ω peak from rocking curve measurements. The root mean square (RMS) roughness was extracted using Gwyddion software over a 1x1 μm$^2$ image for each sample as plotted as a function of $1/t$, where $t$ is the thickness of the films, as shown in figure 2(a). The plot shows a large distribution in the RMS roughness of the Nb thin films strongly dependent on the growth parameters, 1-7 Å for all samples, and 3-7 Å for samples with the thickness in the range of 40~50 nm. For set C and set D with a constant film thickness (50 nm), the RMS roughness decreased from 7 Å to 4 Å as a result of the reduced target voltage; while for set B, the roughness was further reduced to 2 Å with higher target current. The effect of target bias and ion flux on the surface roughness, intermixing interface, and the magnetic properties of free layer in magnetic tunnel junction, and of Ta/Cu thin films were studied previously and revealed similar results [23,24].

Figure 2(b)-(c) show the effect of target bias voltage on the density of grain and on the grain size. For higher target voltage [figure 2(c)], the grains were more elongated compared to the film deposited under lower target voltage [figure 2(b)]. The average grain width for the elongated grains and diameter for the equiaxed grains were summarized in table I. In sample sets A and B, the samples with equiaxed grains are marked with * (table I), whereas all the samples in set C and D had the elongated grains. The grain sizes estimated from AFM images were comparable to the crystallite sizes calculated based on the width of Nb (110) peaks in $2\theta$ scans.

The superconducting transition temperature for all samples were measured and extracted from resistance as a function of temperature curves, and was plotted as a function of $1/t$ as seen in figure 3(a). For the control set A, the dependence of $T_c$ as a function of the film thickness was in good agreement with previous reports of superconducting Nb thin films deposited by other techniques [8,13]. The value of $T_c$, approached to the bulk value as the effect of "surface" on the superconductivity was insignificant when the thickness of the film became much larger than the penetration depth of Nb [25]. The fitting (red line) was fit to a control set A where the deposition conditions were fixed to 400 V and 500 mA of target bias using the following equation by Cooper [26] (assuming $a \ll t$):

$$T_c(t) = T_{c0} \exp(-\frac{2a}{NVt})$$

where $T_c(t)$ is the transition temperature of the film thickness $t$, $T_{c0}$ is the bulk transition temperature, $NV$ is the bulk interaction potential which is 0.32 for niobium [27], and $a$ is the thickness of a surface layer with degraded superconductivity. The fitting yielded $T_{c0}$ of 9.4 ± 0.3 K, which is very close to that of bulk Nb (~9.3 K). The thickness of the surface layer ($a$) was estimated to be 0.75 nm from the fitting. This is in the same order but slightly higher than the values of 0.55-0.58 nm reported in previous work [8,28].

The inset in Figure 3(a) is resistance as a function of temperature of one sample near $T_c$ showing the heat-up and cool-down curves. The superconducting transition width $\Delta T_c$ for all samples reported in this work was 0.1 K or less. The small hysteresis implied that the superconductivity is very homogenous through the entire film, and in fact the value observed here was much smaller than samples deposited by other methods such as dc sputtering or MBE [7,13].

It was observed that under the same growth conditions (set A), as the film thickness increased, the FWHM (ω) of Nb peak was reduced thanks to the improved crystallinity [figure 3(b)], but also it resulted in a rougher surface [figure 3(d)]. Thicker films resulted in the properties that is approaching bulk value for both higher $T_c$ [figure 3(a)] and better crystallinity which is comparable to previous results [8]. The grain width obtained from AFM analysis as a function of thickness is shown in figure 3(c), where the thinnest sample A1 (circled) had equiaxed grains different than the thicker samples with elongated grains. The width of elongated grains in set A increased as the thickness increased, raising the $T_c$, which agrees well with previous reported results [13,29].

Figure 4 shows the result of set B, where the deposition rate, FWHM, $T_c$ and roughness are plotted as a function of the target current, when the samples were deposited at a constant target voltage of 400 V. Both $T_c$ and roughness decreased as the target current increased. The films deposited at higher target current (green circle) were thicker than the other two, yet still had much smoother surface; however, they showed poor crystallinity, much larger FWHM as shown in figure 4(b). These results contradicted the thickness dependent results obtained from set A where the thicker film had better crystallinity, higher $T_c$, and rougher surface. It suggests that the reduced $T_c$ was the result of poorer crystallinity for samples B3 and B4. Moreover the two samples in the circle (B3, B4) also had equiaxed grains instead of the elongated grains that the better crystallinity samples had. As the target current increased, the deposition rate was also higher [figure 4(a)], however the crystallinity was not monotonically reduced. Instead, the film deposited at the intermediate rate of 0.8 Å/s showed the best crystallinity. The result suggests that higher target current not only reduced the surface roughness, but also improved the crystallinity of the Nb. However, eventually the higher ion flux and energy led to bombardment damages to deteriorate the crystallinity, thus degraded the superconductivity of the Nb thin film by introducing defects as shown in samples B3 and B4.

With the same target bias voltage, raising the target current required higher anode current (as presented in table I). With fixed Ar flow rate through the anode, increasing the anode current also raised the anode voltage, hence yielding higher ion energy. Cuomo *et. al.* found that the low-energy ion flux can modify the stress of the Nb films [10], which could be related to the modification of microstructure and grain boundaries of the films. Cuomo *et. al.* also reported that these films had higher tensile stress (in-plane) and better conductivity as the ion flux was increased for the sample set deposited at low temperature. This was opposite of what was observed here. For set B, as the ion flux was increased, the lattice spacing $d_{110}$ is extended and become more tensile as shown in table I. Assuming a conservation of unit cell volume, the in-plane lattice is more compressive, as ion flux was increased. This is likely because of the rising ion energy as the ion flux was increased for the BTIBD technique. As aforementioned, the $T_c$ was decreasing when the anode current used during the deposition was higher (set B).

Comparing the results from sample sets A and B, it was obvious that the crystallinity of Nb thin film strongly depended on the film thickness and the deposition rate (target current). To discern the correlations among grain shape, thickness, and crystallinity, set C and D were deposited with low to intermediate target current, where the deposition rate was controlled to be 0.5 Å/s and 0.8 Å/s respectively. The thickness of the films in both sets were fixed at 50 nm, while the target voltage is varied from 400 to 900 V. $T_c$, grain width, and roughness are plotted as function of target voltage as shown in figures 5(a)-(c), respectively. The low target bias encouraged the growth of Nb grains, and also led to the smoother surface, and higher $T_c$. The grain width as a function of target voltage showed the same trend as the crystallite size for both sets (table I). While the FWHM of Nb peak was roughly constant within the sets, implying that the crystallinity of Nb films of given thickness was insensitive to the growth conditions explored in this study. A slight change in the FWHM was likely from the small thickness variation (table I).

As aforementioned, the width of elongated grains in set A increased as the thickness increased, while the two samples from set B in the green circle presented in figure 4(d) had the equiaxed grains instead of the elongated grains. In sets C and D, all samples had the elongated grains with decreasing grain width as the target voltage increased. For these two sets (C and D), the energy flux associated with the arriving adatoms on the surface was increased because of the high ion kinetic energy as the target voltage increased, and the effect on the grain size could be the result of the changes in ion energy and ion flux. The evolution of morphology observed in this study was consistent with the extended structure zone diagram proposed by Anders [30]. The Nb films with equiaxed grains are in the zone for "nanocrystalline with preferred orientation", and those with elongated grains are in the zone for "recrystallized grains". The crystallinity of Nb appeared to be independent if the different microstructures and morphologies, which may be characteristics of Nb crystal. Nonetheless, it is evident that the energy and flux of ions (target bias and current) played important roles in controlling the morphology and the surface roughness of Nb films, as well as the evolution of microstructure.

Comparing each pair of samples in set C and set D deposited at the same target voltage, the 0.8 Å/s sample set (set D) provided better crystallinity of Nb, and smoother surface which was likely a result from the higher ion flux and energy, while the $T_c$ was comparable. The width of elongated grains was strongly dependent on the target voltage, even though the anode current was also changing within set C and set D. This is because the correlation of the grain width and the anode current in these two sets contradicted that of set B where the target voltage was constant. The higher anode current in set B yielded narrower grains, while the higher anode current in set C and in set D yielded wider grains as seen in table I. In other words, the decreasing grain size in set C and D resulted predominantly from the change in the target voltage rather than the change in the anode current.

$T_c$ was plotted as a function of target current (anode current) of set C and D in figure 6(a). This result implies that $T_c$ was slightly improved by reduced target bias voltage. There appears to be a weak relationship between $T_c$ and grain width in set C and set D as shown in figure 6(b). Larger grain size resulted in higher $T_c$, which agrees well with previous reports [13,29].

To summarize the effects of target current and bias, it appears that the surface roughness is reduced by increasing target current, while the $T_c$ and crystallinity is dependent of target voltage. Using low target bias while maintaining intermediate target current during the deposition can reduce the surface roughness while improving the critical temperature for superconductivity in Nb thin films. It is worth noting that the crystallinity was not degraded with this approach.

## 4. CONCLUSION

In summary, we explored various growth parameters to prepare Nb thin film samples to control the crystallinity of Nb and the surface roughness. Too high target current (> 500 mA at 400 V bias) during the deposition degraded the Nb crystallinity, and subsequently reduced the critical temperature for superconductivity. It was found that the surface morphology, i.e. grain size was modified by varying the target bias. The low target bias (400 V) yielded smoother Nb films, larger grains, and higher $T_c$. To reduce the thickness of the proximity layer of Nb, preparing a smoother Nb surface was achieved by lowering the target voltage while maintaining intermediate target current (~ 500 mA). With the target bias and current of 400 V and 500 mA, we obtained smooth Nb surface (RMS roughness of 4.08 Å) without degrading the crystallinity and the superconductivity of the Nb thin film. The preparation of such smooth Nb with adequate superconductivity can serve as a base electrode for the in-situ magnetic layer or insulating layer for superconducting electronic devices.

ACKNOWLEDGEMENT

The authors are grateful to the support from Northrop Grumman Corporation and Intelligence Advanced Research Projects Activity (IARPA).

**Table I.** Growth conditions and characterization of Nb thin film samples (* denote samples with equiaxed grains), where $V_T$ is target voltage, $I_T$ is target current, and $I_A$ is anode current.

|    | $V_T$ (V) | $I_T$ (mA) | $I_A$ (A) | rate (Å/s) | $t$ (nm) | roughness (Å) | $d_{110}$ (Å) | Crystal. (nm) | Grain (nm) | FWHM ω (deg) | $T_c$ (K) |
|----|-----------|------------|-----------|------------|----------|---------------|---------------|---------------|------------|--------------|-----------|
| A1 | 400 | 500 | 7.0 | 0.91 | 11.7 | 1.39 | 3.3544 | 10.50 | *14.7** | 10.55 | 6.20 |
| A2 | 400 | 500 | 7.0 | 0.83 | 21.2 | 2.47 | 3.3536 | 10.76 | 10.8 | 8.48 | 7.80 |
| A3 | 400 | 500 | 7.0 | 0.81 | 31.0 | 2.57 | 3.3509 | 11.35 | 11.4 | 6.88 | 8.10 |
| A4 | 400 | 500 | 7.0 | 0.79 | 40.4 | 2.88 | 3.3454 | 12.57 | 14.3 | 6.30 | 8.35 |
| A5 | 400 | 500 | 7.0 | 0.90 | 92.5 | 6.09 | 3.3390 | 12.88 | 15.7 | 5.10 | 8.85 |
| B1 | 400 | 300 | 4.5 | 0.45 | 45.0 | 4.34 | 3.3372 | 12.45 | 18.3 | 7.59 | 8.80 |
| B2 | 400 | 500 | 7.0 | 0.78 | 47.1 | 4.08 | 3.3373 | 12.02 | 14.8 | 6.39 | 8.75 |
| B3 | 400 | 530 | 7.4 | 0.83 | 74.5 | 2.40 | 3.3498 | 19.53 | *22.3** | 10.20 | 8.30 |
| B4 | 400 | 645 | 8.5 | 1.01 | 66.5 | 2.07 | 3.3567 | 11.64 | *21.2** | 10.78 | 7.35 |
| C1 | 400 | 300 | 4.5 | 0.45 | 45.0 | 4.34 | 3.3372 | 12.45 | 18.3 | 7.59 | 8.80 |
| C2 | 600 | 250 | 3.9 | 0.48 | 48.2 | 5.34 | 3.3410 | 12.40 | 15.6 | 7.39 | 8.75 |
| C3 | 800 | 210 | 3.3 | 0.48 | 48.3 | 6.39 | 3.3451 | 11.83 | 13.9 | 7.32 | 8.65 |
| C4 | 900 | 200 | 3.2 | 0.49 | 49.3 | 6.61 | 3.3498 | 10.64 | 8.7 | 7.22 | 8.55 |
| D1 | 400 | 500 | 7.0 | 0.78 | 47.1 | 4.08 | 3.3373 | 12.02 | 14.8 | 6.39 | 8.75 |
| D2 | 600 | 432 | 6.0 | 0.84 | 50.7 | 4.68 | 3.3464 | 10.21 | 13.4 | 6.13 | 8.70 |
| D3 | 800 | 363 | 5.1 | 0.84 | 50.5 | 4.95 | 3.3455 | 10.04 | 11.9 | 6.16 | 8.65 |
| D4 | 900 | 339 | 4.9 | 0.82 | 49.6 | 6.11 | 3.3483 | 9.99 | 10.2 | 6.22 | 8.55 |

**FIGURE CAPTIONS**

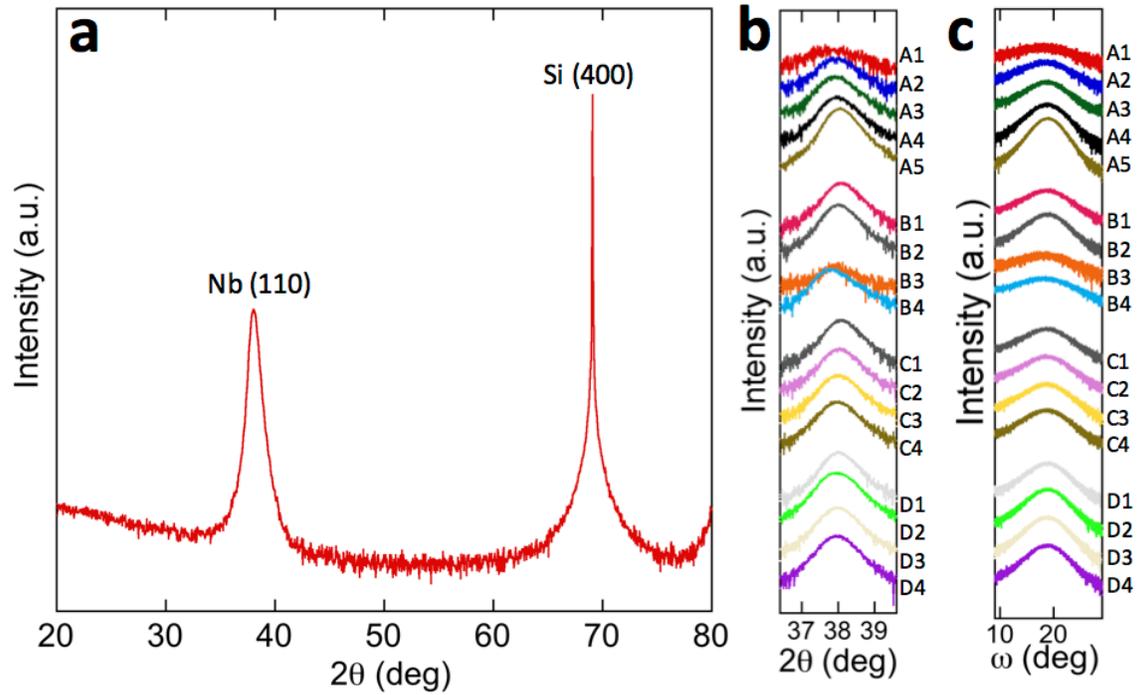

Fig. 1 (a) an example of XRD 2θ scan showing Nb (110) and Si (400) peaks of sample A5; (b) 2θ scan showing Nb (110) for all samples; (c) rocking curve measurement on Nb (110) for all samples.

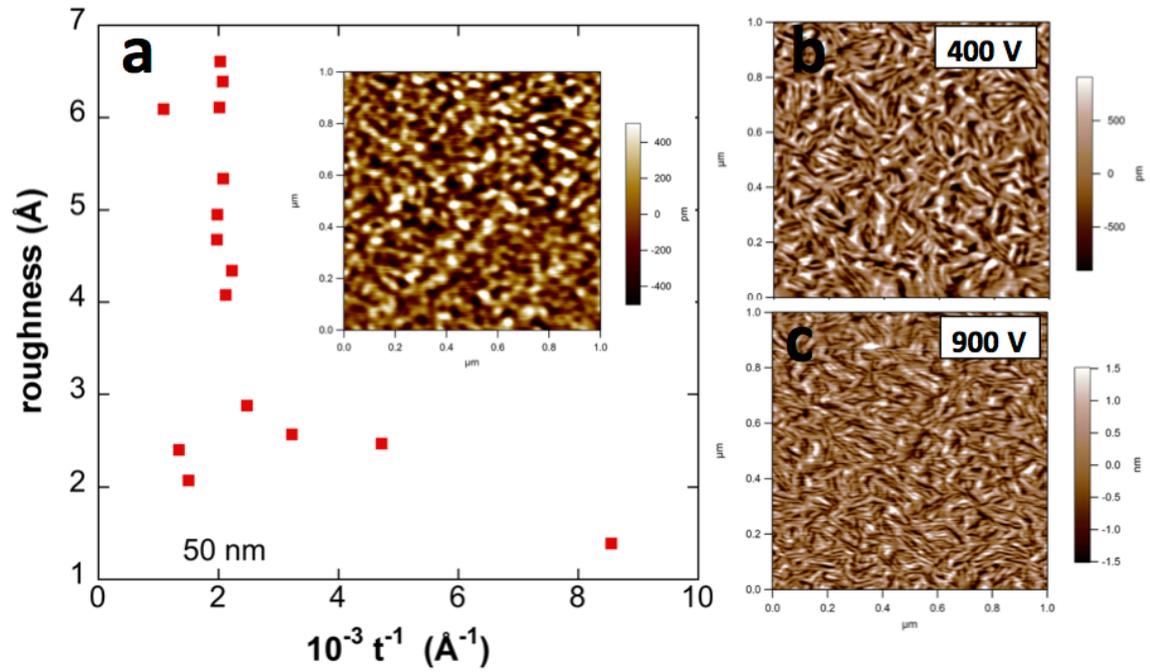

Fig. 2 (a) RMS roughness samples as functions of $1/t$, where $t$ is the thickness of the films, inset is a 1 x 1 μm² area AFM image showing surface with equiaxed grains; AFM images showing surface with elongated grains of the sample deposited at (b) 400 V, and (c) 900 V target bias.

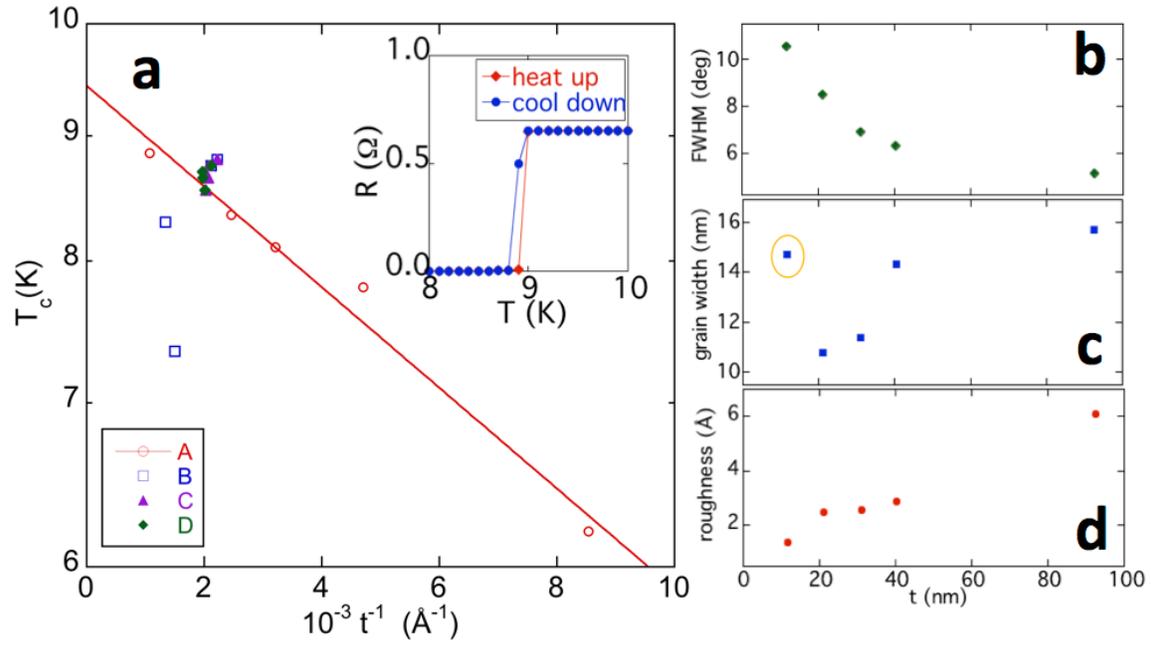

Fig. 3 (a) superconducting transition temperature of all samples as functions of 1/*t*, where *t* is the thickness of the films, the red line show the fitted curve of sample set A; inset shows R vs. T curve near $T_c$ of one sample; (b) FWHM of ω scan, (c) grain width, and (d) RMS roughness as functions of thickness for sample set A. The sample in yellow circle has equiaxed grains.

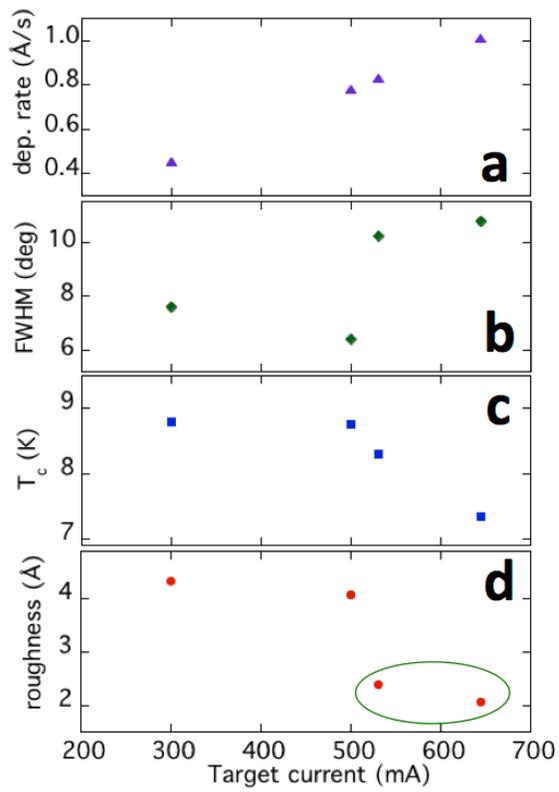

Fig. 4 Effect of target current on (a) deposition rate, (b) FWHM of ω scan, (c) $T_c$, (d) RMS roughness for sample set B deposited at fixed 400 V target bias. The crystal grains are equiaxed in the two samples in the green circle.

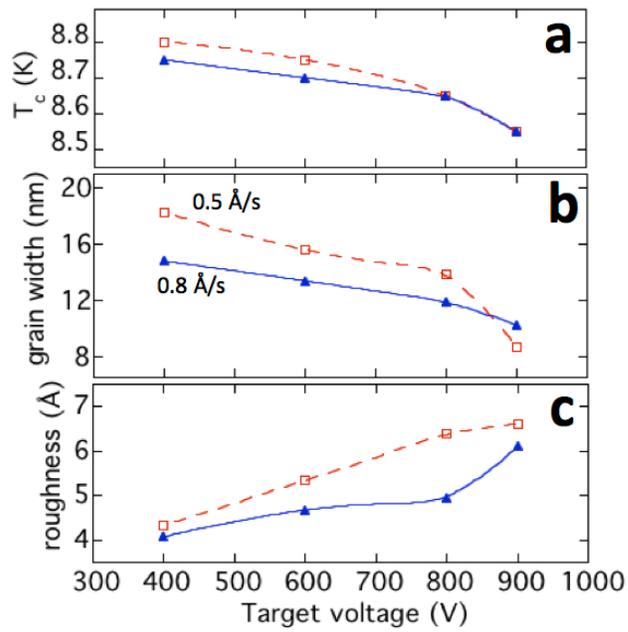

Fig. 5 Effect of target bias on (a) $T_c$, (b) grain width, and (c) RMS roughness as functions of target bias voltage for sample set C and D. Dashed line represents set C (0.5 Å/s), while solid line represents set D (0.8 Å/s).

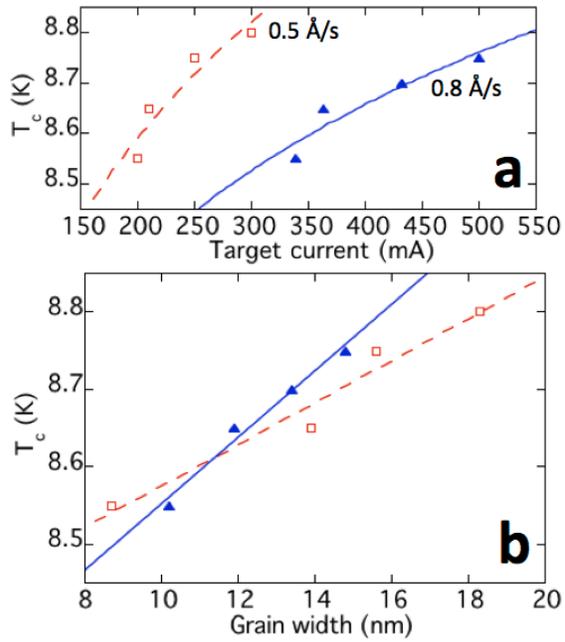

Fig. 6 (a) $T_c$ as a function of target current for sample set C and set D, contradicting the result from set B; (b) $T_c$ as a function of grain width for sample set C and D. Dashed line represents set C (0.5 Å/s), while solid line represents set D (0.8 Å/s).